\documentclass[preprintnumbers,superscriptaddress,pra]{revtex4}
\usepackage{amssymb}
\usepackage{amsmath}
\usepackage{epsfig}
\usepackage{graphicx}
\usepackage{color}

\input{tcilatex}
\begin{document}

\title{Phonon number fluctuations in Debye model of solid}
\author{Q. Chen}
\affiliation{School for Theoretical Physics, School of Physics and Electronics, Hunan
University, Changsha 410082, China}
\author{Q. H. Liu}
\email{quanhuiliu@gmail.com}
\affiliation{School for Theoretical Physics, School of Physics and Electronics, Hunan
University, Changsha 410082, China}
\affiliation{Synergetic Innovation Center for Quantum Effects and Applications (SICQEA),
Hunan Normal University,Changsha 410081, China}
\date{\today}

\begin{abstract}
The phonon number fluctuations in the Debye model of solid are calculated
and are demonstrated to be proportional to the temperature cubed at low
temperature, similar to the celebrated Debye's law of the heat capacity. For
a fixed number of atoms, the relative fluctuations approach to infinity as
the temperatture decreases to zero, and the proper definition of temperature
needs more and more numbers of atoms at lower and lower temperatures,
compatible with the third law on the unattainability of absolute zero
temperature.
\end{abstract}

\maketitle

\section{Introduction}

Fluctuations are ubiquitous in the Universe and the statistical mechanics is
powerful to understand them. The Debye model of solid was the first
successful model used to describe the heat capacity of a material at low
temperature, via introduction of the concept phonon into physics. However,
the phonon number fluctuations in the model are not yet known, and we will
demonstrate that they fall off as $T^{3}$ at low temperature $T\rightarrow0$%
, similar to the decrease of the heat capacity of solid at low temperature.

In the Debye model of the solid, the thermal properties are determined by
the solid lattice vibrations. The vibrational frequencies form a continuous
spectrum with a cut off at an upper limit $\omega_{D}$ such that the total
number of normal modes of vibration is $3N$ of which $N$ are the number of
the solid atoms which can harmonically displaced from lattices. The Debye
spectrum $g(\omega)$ or density of states in frequency interval $%
\omega\rightarrow \omega+d\omega$ is, \cite{text1,text2,text3,text4}%
\begin{equation}
g(\omega)=\left\{ 
\begin{array}{cc}
\frac{9N}{\omega_{D}^{3}}\omega^{2}, & \omega\leq\omega_{D} \\ 
0 & \omega>\omega_{D}%
\end{array}
\right\} ,\text{ \ }\omega_{D}=\left( 6\pi^{2}\right) ^{1/3}c\left( \frac{N}{%
V}\right) ^{1/3},   \label{DOS}
\end{equation}
where symbols $c$ and $V$ denote the effective speed of sound within the
solid and its volume, respectively. Two quantities $g(\omega)$ and $%
\omega_{D}$ are related by the requirement that total number of normal modes
of vibration $3N$, 
\begin{equation}
\int_{0}^{\omega_{D}}g(\omega)d\omega=3N.   \label{Fcut}
\end{equation}
Assuming that there are $n_{i}$ phonons in the $i$th frequency $\omega_{i}$
whose unit energy quantum is $\hbar\omega_{i}$, we have the energy of the
state $\{n_{i}\}$ in which there are $n_{i}$ phonons of the $i$th mode is,%
\begin{equation}
E\{n_{i}\}=\sum_{i=1}^{3N}n_{i}\hbar\omega_{i}.   \label{AllEnergy}
\end{equation}
The partition function is, \cite{text1,text2,text3,text4}%
\begin{equation}
Q=\sum_{\{n_{i}\}}e^{-\beta
E\{n_{i}\}}=\prod_{i=1}^{3N}\sum_{n_{i}=0}^{\infty}e^{-n_{i}\beta\hbar%
\omega_{i}}=\prod_{i=1}^{3N}\frac{1}{1-e^{-\beta\hbar\omega_{i}}}. 
\label{PartF}
\end{equation}
where $\beta=1/kT$, and $\hbar$ and $k$ are, respectively, the Planck's
constant and the Boltzmann's constant. The average number $\left\langle
n_{i}\right\rangle $ of phonon of energy quantum $\hbar\omega_{i}$ is,%
\begin{equation}
\left\langle n_{i}\right\rangle \equiv\frac{\sum_{\{n_{i}\}}n_{i}e^{-\beta
E\{n_{i}\}}}{Q}=\frac{\sum_{n_{i}=0}^{\infty}n_{i}e^{-n_{i}\beta\hbar
\omega_{i}}}{\sum_{n_{i}=0}^{\infty}e^{-n_{i}\beta\hbar\omega_{i}}}=\frac {1%
}{e^{\beta\hbar\omega_{i}}-1}=-kT\frac{\partial}{\partial\left( \beta
\hbar\omega_{i}\right) }\ln Q.   \label{M1}
\end{equation}
The internal energy is,%
\begin{equation}
U=\sum_{i=1}^{3N}\left\langle n_{i}\right\rangle \hbar\omega_{i}=\sum
_{i=1}^{3N}\frac{\hbar\omega_{i}}{e^{\beta\hbar\omega_{i}}-1}%
=\int_{0}^{\omega_{D}}\frac{\hbar\omega}{e^{\beta\hbar\omega}-1}%
g(\omega)d\omega .   \label{U}
\end{equation}
A more detailed discussion of this integral can be given shortly. However,
the internal energy can be greatly simplified in limits of high and low
temperatures, respectively, \cite{text1,text2,text3,text4}%
\begin{equation}
U\simeq3NkT\left\{ 
\begin{array}{cc}
1, & T\gg T_{D} \\ 
\frac{\pi^{4}}{5}(\frac{T}{T_{D}})^{3}, & T\ll T_{D}%
\end{array}
\right. ,   \label{2Limits}
\end{equation}
where $T_{D}$ is the Debye temperature defined by,%
\begin{equation}
kT_{D}=\hbar\omega_{D}.   \label{DebyeT}
\end{equation}
The values of the heat capacity $C=\partial U/\partial T$ in two opposite
limits are,%
\begin{equation}
C\simeq3Nk\left\{ 
\begin{array}{cc}
1, & T\gg T_{D} \\ 
\frac{4\pi^{4}}{5}(\frac{T}{T_{D}})^{3}, & T\ll T_{D}%
\end{array}
\right. .   \label{HeatC}
\end{equation}
This equation is significant for it in high temperature limit gives the
Dulong--Petit law, and in low temperature limit yields the $T^{3}$ law.

The number fluctuations of the phonon is defined by,%
\begin{equation}
\left\langle n^{2}\right\rangle =\left\langle
\sum_{r}n_{r}\sum_{s}n_{s}\right\rangle ,\text{ and }\left\langle
n\right\rangle =\sum_{i=1}^{3N}\left\langle n_{i}\right\rangle .
\end{equation}
How to calculate it will be given in section II. In section III, brief
concluding remarks are given.

The Debye integral $D_{n}(x)$ \cite{text5} is elementary in our analysis, 
\begin{equation}
D_{n}(x)\equiv\int_{0}^{x}\frac{t^{n}}{e^{t}-1}dt,   \label{DebyeI}
\end{equation}
which has simple expressions in limits of large and small $x$ with $\zeta$
denoting the Riemann zeta function 
\begin{equation}
D_{n}(x)\simeq\left\{ 
\begin{array}{cc}
n!\zeta(n+1)-x^{n}e^{-x}+O(x^{n}e^{-2x}), & x\rightarrow\infty \\ 
x^{n}/n+O(x^{n+1}), & x\rightarrow0%
\end{array}
\right. .   \label{Dn}
\end{equation}

\section{The mean numbers and the number fluctuations}

The mean numbers for phonons in the Debye model of solid are from (\ref{M1}),%
\begin{align}
\left\langle n\right\rangle & =\sum_{i=1}^{3N}\left\langle n_{i}\right\rangle
\notag \\
& =\int_{0}^{\omega_{D}}\frac{1}{e^{\beta\hbar\omega}-1}g(\omega )d\omega 
\notag \\
& =\frac{9N}{\omega_{D}^{3}}\int_{0}^{\omega_{D}}\frac{\omega^{2}}{%
e^{\beta\hbar\omega}-1}d\omega.   \label{Mean}
\end{align}
Performing a variable transformation $t=\beta\hbar\omega$ and defining, 
\begin{equation}
t_{D}\equiv\frac{\hbar\omega_{D}}{kT}=\frac{T_{D}}{T},   \label{tD}
\end{equation}
the mean numbers $\left\langle n\right\rangle $ (\ref{Mean}) becomes,%
\begin{equation}
\left\langle n\right\rangle =\frac{9N}{\omega_{D}^{3}}\left( \frac{kT}{\hbar 
}\right) ^{3}\int_{0}^{t_{D}}\frac{t^{2}}{e^{t}-1}%
dt=9Nt_{D}^{-3}D_{2}(t_{D}).   \label{M-D2}
\end{equation}
It is thus in the opposite limits of temperature intervals from (\ref{Dn}),%
\begin{equation}
\left\langle n\right\rangle \simeq9N\left( \frac{T}{T_{D}}\right)
^{3}\left\{ 
\begin{array}{cc}
\frac{1}{2}\left( \frac{T_{D}}{T}\right) ^{2}, & T\gg T_{D} \\ 
2\zeta(3), & T\ll T_{D}%
\end{array}
\right. =9N\left\{ 
\begin{array}{cc}
\frac{1}{2}\frac{T}{T_{D}}, & T\gg T_{D} \\ 
2\zeta(3)\left( \frac{T}{T_{D}}\right) ^{3}, & T\ll T_{D}%
\end{array}
\right. .   \label{nMean}
\end{equation}

Before computing the particle number fluctuations, we need to deal
statistical correlation $\left\langle n_{r}n_{s}\right\rangle $ of the
particle number. If $r\neq s$, we have, 
\begin{align}
\left\langle n_{r}n_{s}\right\rangle & \equiv\frac{\sum_{\{n_{i}%
\}}n_{r}n_{s}e^{-\beta E\{n_{i}\}}}{Q}  \notag \\
& =\frac{\sum_{n_{r}=0}^{\infty}\sum_{n_{s}=0}^{\infty}n_{r}n_{s}e^{-\left(
n_{r}\beta\hbar\omega_{r}+n_{s}\beta\hbar\omega_{s}\right) }}{%
\sum_{n_{r}=0}^{\infty}\sum_{n_{s}=0}^{\infty}e^{-\left(
n_{r}\beta\hbar\omega_{r}+n_{s}\beta\hbar\omega_{s}\right) }}  \notag \\
& =\frac{\sum_{n_{r}=0}^{\infty}n_{r}e^{-n_{r}\beta\hbar\omega_{r}}}{%
\sum_{n_{r}=0}^{\infty}e^{-n_{r}\beta\hbar\omega_{r}}}\frac{%
\sum_{n_{s}=0}^{\infty}n_{s}e^{-n_{s}\beta\hbar\omega_{s}}}{%
\sum_{n_{s}=0}^{\infty }e^{-n_{s}\beta\hbar\omega_{s}}}  \notag \\
& =\left\langle n_{r}\right\rangle \left\langle n_{s}\right\rangle . 
\label{correlation}
\end{align}
The minus of the particle number fluctuations in the $r$th vibration mode
are,%
\begin{equation}
-\left( \left\langle n_{r}^{2}\right\rangle -\left\langle n_{r}\right\rangle
^{2}\right) =\frac{\partial}{\partial\left( \beta\hbar\omega_{r}\right) }%
\frac{\sum_{n_{r}=0}^{\infty}n_{r}e^{-n_{r}\beta\hbar\omega_{r}}}{\sum
_{n_{r}=0}^{\infty}e^{-n_{r}\beta\hbar\omega_{r}}}.   \label{rem1}
\end{equation}
The right-handed side of this equation becomes from (\ref{M1}),%
\begin{align}
\frac{\partial}{\partial\beta\hbar\omega_{r}}\frac{1}{e^{\beta\hbar%
\omega_{i}}-1} & =-\frac{e^{\beta\hbar\omega_{i}}}{\left(
e^{\beta\hbar\omega_{i}}-1\right) ^{2}}  \notag \\
& =-\frac{e^{\beta\hbar\omega_{i}}-1+1}{\left(
e^{\beta\hbar\omega_{i}}-1\right) ^{2}}  \notag \\
& =-\left( \left\langle n_{r}\right\rangle ^{2}+\left\langle
n_{r}\right\rangle \right) .   \label{rem2}
\end{align}
Combining two results (\ref{rem1})-(\ref{rem2}), we reach a remarkable
result,%
\begin{equation}
\left\langle n_{r}^{2}\right\rangle -\left\langle n_{r}\right\rangle
^{2}=\left\langle n_{r}\right\rangle ^{2}+\left\langle n_{r}\right\rangle .
\end{equation}
The fluctuations in phonon number are,%
\begin{align}
\left\langle \Delta n^{2}\right\rangle & \equiv\left\langle
n^{2}\right\rangle -\left\langle n\right\rangle ^{2}  \notag \\
& =\left\langle \sum_{r}n_{r}\sum_{s}n_{s}\right\rangle -\left\langle \sum
_{r}n_{r}\right\rangle \left\langle \sum_{s}n_{s}\right\rangle  \notag \\
& =\sum_{r=1}^{3N}\left( \left\langle n_{r}^{2}\right\rangle -\left\langle
n_{r}\right\rangle ^{2}\right)  \notag \\
& =\sum_{r=1}^{3N}\left( \left\langle n_{r}\right\rangle ^{2}+\left\langle
n_{r}\right\rangle \right) ,
\end{align}
which can be transformed into an integral,%
\begin{align}
\left\langle \Delta n^{2}\right\rangle & =\frac{9N}{\left( \omega
_{D}\right) ^{3}}\int_{0}^{\omega_{D}}\frac{\omega^{2}e^{\beta\hbar\omega}}{%
\left( e^{\beta\hbar\omega}-1\right) ^{2}}d\omega  \notag \\
& =9Nt_{D}^{-3}\int_{0}^{t_{D}}\frac{t^{2}e^{t}}{\left( e^{t}-1\right) ^{2}}%
dt  \notag \\
& =9Nt_{D}^{-3}\left( 2D_{1}(t_{D})-\frac{t_{D}^{2}}{e^{t_{D}}-1}\right) . 
\label{n-f}
\end{align}
In limits of high and low $T$, Eq. (\ref{n-f}) gives with $\zeta(2)=\pi^{2}/6
$, 
\begin{equation}
\left\langle \Delta n^{2}\right\rangle \simeq9N\left( \frac{T}{T_{D}}\right)
^{3}\left\{ 
\begin{array}{cc}
\frac{T_{D}}{T}, & T\gg T_{D} \\ 
\pi^{2}/3, & T\ll T_{D}%
\end{array}
\right. .
\end{equation}
The relative fluctuations are,%
\begin{align}
\frac{\left\langle \Delta n^{2}\right\rangle }{\left\langle n\right\rangle
^{2}} & \simeq\frac{1}{9N\left( \frac{T}{T_{D}}\right) ^{3}}\left\{ 
\begin{array}{cc}
4\left( \frac{T}{T_{D}}\right) ^{3}, & T\gg T_{D} \\ 
\frac{\pi^{2}}{3}\frac{1}{4\zeta(3)^{2}}, & T\ll T_{D}%
\end{array}
\right.  \notag \\
& =\frac{1}{9N}\left\{ 
\begin{array}{cc}
4, & T\gg T_{D} \\ 
0.569\ \left( \frac{T_{D}}{T}\right) ^{3}, & T\ll T_{D}%
\end{array}
\right. .   \label{relativeF}
\end{align}
In addition we have from (\ref{n-f}) and (\ref{M-D2}), 
\begin{align}
\frac{\left\langle \Delta n^{2}\right\rangle }{\left\langle n\right\rangle }
& =\frac{2D_{1}(t_{D})-\frac{t_{D}^{2}}{e^{t_{D}}-1}}{D_{2}(t_{D})}  \notag
\\
& \simeq\left\{ 
\begin{array}{cc}
2T/T_{D}, & T\gg T_{D} \\ 
\zeta(2)/\zeta(3), & T\ll T_{D}%
\end{array}
\right. .
\end{align}
It is interesting to note that with a given large number of atoms $N$, the
relative fluctuation is not automatically less than $1$ at low temperature.
I.e, the following equation would be violated,%
\begin{equation}
\frac{\left\langle \Delta n^{2}\right\rangle }{\left\langle n\right\rangle
^{2}}\simeq\frac{0.063}{N}\left( \frac{T_{D}}{T}\right) ^{3}\leq1.
\end{equation}
The self-consistence of the statistical mechanics implies a requirement upon
the temperature,%
\begin{equation}
T\gtrsim0.398T_{D}N^{-1/3}.
\end{equation}
In other words, when temperature approaches to zero Kelvin, the
thermodynamic limit requires a huge number of particles $N\rightarrow\infty$
otherwise the low temperatures can not be properly defined, \cite{Liu}
compatible with the third law on the unattainability of absolute zero
temperature.

\section{Remarks}

The mean numbers and the number fluctuations of the phonon in the Debye
model are explicitly calculated. The numbers and fluctuations are
demonstrated to be proportional to the temperature cubed at low temperature,
similar to the celebrated Debye's law of the heat capacity. In addition, the
relative fluctuations diverge for a fixed number of atoms, implying that the
proper definition of temperature needs more and more numbers of atoms at
lower and lower temperatures.

\begin{acknowledgments}
This work is financially supported by National Natural Science Foundation of
China under Grant No. 11675051.
\end{acknowledgments}

\end{document}